\documentclass{LMCS}

\usepackage{enumerate}
\usepackage{hyperref}
\usepackage{amsmath}
\theoremstyle{plain}
\newtheorem{lemma}[thm]{Lemma}
\newtheorem{example}[thm]{Example}

\newcommand\ignore[1]{}

\newcommand\iffullpaper[1]{}
\newcommand{\comment}[1]{}
\newcommand\subterms[1]{{{\tt Subterms}(}{#1}{)}}
\newcommand\Pos{{\tt Pos}}
\newcommand{\X}{{\mathcal X}}

\def\Tau{{\mathcal T}}
\def\Vars{{\mathcal V}}
\newcommand{\tor}{\to_R}

\def\doi{6 (3:8) 2010}
\lmcsheading%
{\doi}
{1--20}
{}
{}
{Dec.~\phantom04, 2009}
{Aug.~25, 2010}
{}   

\begin{document}

\title[Termination of Rewriting]{Termination of Rewriting with Right-Flat Rules Modulo Permutative Theories\rsuper*}

\author[L.~Bargu\~n\'o]{Luis Bargu\~n\'o\rsuper a}	%
\address{{\lsuper{a,b,c}}Universitat Polit\`ecnica de Catalunya, Jordi Girona 1, Barcelona, Spain}
\email{luisbargu@gmail.com, ggodoy@lsi.upc.edu, eduard.hl@gmail.com}  %
\thanks{{\lsuper a}Supported by Spanish Ministry of Education and Science by the
FORMALISM project (TIN2007-66523).}

\author[G.~Godoy]{Guillem Godoy\rsuper b}	%
\address{\vskip-6 pt}
\thanks{{\lsuper b}Supported by Spanish Ministry of Education and Science by the
FORMALISM project (TIN2007-66523) and the
LOGICTOOLS-2 project (TIN2007-68093-C02-01).}

\author[E.~Huntingford]{Eduard Huntingford}	%
\address{\vskip-6 pt}
%
%
%

\author[A.~Tiwari]{Ashish Tiwari\rsuper c}	%
\address{{\lsuper c}SRI International, Menlo Park, CA 94205}	%
\email{tiwari@csl.sri.com}  %
\thanks{{\lsuper c}Supported in part by the National Science Foundation under grants
CNS-0720721 and CSR-0917398.}	%

\titlecomment{{\lsuper*}A preliminary version~\cite{DBLP:conf/rta/GodoyHT07}
containing some of the results
appeared in the Proceedings of the 18th International Conference
on Rewriting Techniques and Applications, RTA 2007.}

\begin{abstract}
\noindent 
We present decidability results for termination of classes
of term rewriting systems modulo permutative theories.
Termination and innermost termination modulo permutative theories
are shown to be decidable for term rewrite systems (TRS) whose right-hand
side terms are restricted to be shallow (variables
occur at depth at most one) and linear (each variable occurs at most once).
Innermost termination modulo permutative theories is also shown to be decidable
for shallow TRS. 
We first show that a shallow TRS can be transformed
into a flat (only variables and constants occur at depth one) 
TRS while preserving termination and innermost
termination.
The decidability results are then proved by showing that 
(a) for right-flat right-linear (flat) TRS, 
non-termination (respectively, innermost non-termination) implies
non-termination starting from flat terms, and
(b) %
for right-flat TRS,
the existence of non-terminating derivations
starting from a given term is decidable.
On the negative side, we show PSPACE-hardness of termination and innermost
termination for shallow right-linear TRS,
and undecidability of termination
for flat TRS.%
\end{abstract}

\maketitle

\section{Introduction}

\noindent Termination is an important property of
computing systems and it has
generated significant renewed interest in 
recent years.  
There has been progress on both the theoretical
and practical aspects of proving
termination of many different computing 
paradigms - such as 
term rewrite systems (TRS), 
functional programs, and 
imperative programs.
Innermost termination refers to termination of rewriting
restricted to the innermost strategy, which forces
the complete evaluation of all the subterms before
a rule is applied at any position. It
corresponds to the ``call by value'' computation of programming languages.
A typical example of a TRS that is innermost terminating
but not terminating is the following~\cite{Toy87b}:
$$
\{f(0,1,x)\rightarrow f(x,x,x),\;c\rightarrow 0,\;c\rightarrow 1\}
$$
The non-terminating derivation
$$
\underline{f(0,1,c)}\rightarrow f(\underline{c},\underline{c},c)
\rightarrow^2 f(0,1,c)\rightarrow \cdots
$$
is not possible with innermost rewriting, since $c$ has to
be normalized before a rule can be applied at the root position 
to reduce $f(0,1,c)$.

Often, a term rewrite system contains rules
that are trivially non-terminating (like commutativity:
$f(x,y)\to f(y,x)$) and one desires to ensure a weaker
notion than termination, namely 
termination of a term rewrite system $R$ modulo a theory $E$
(for example, when $E=\{f(x,y)\to f(y,x)\}$).
Althought $R\cup E$ could be non-terminating, in some
cases the important question is to determine if there is
a derivation with $R\cup E$ that has infinitely many rewrite
steps with rules in $R$.

While termination is
undecidable for general TRS
and string rewrite systems~\cite{HuetLankford78},
several subclasses with decidable termination
problem have been identified.  
Termination is decidable for ground TRS~\cite{HuetLankford78};
in fact, in polynomial time~\cite{Plaisted93:RTA}.
Termination is decidable for right-ground
TRS~\cite{Dershowitz81:ICALP} and also for the more general class
that also has collapsing (right-variable) rules~\cite{GodoyTiwari04:IJCAR}. 
Later, it was shown that termination is decidable for
TRS that contain any combination of right-ground, collapsing, and
shallow right-linear rewrite rules~\cite{GodoyTiwari05:CADE}.
There are further known decidability results about
shallow left-linear and shallow right-linear TRS~\cite{Sakai06}.

This paper focuses on termination and innermost termination of TRS
for rewriting modulo permutative theories.
Here we extend the results of our conference
paper~\cite{DBLP:conf/rta/GodoyHT07} by generalizing from
plain rewriting to rewriting modulo permutative theories.
Moreover, 
we provide extended proofs of our earlier results, and
a new PSPACE-hardness result.

The main contributions of the paper are as follows:
\\ (1)
In Section~\ref{sec-right-flat}, we prove that
termination {\em starting from a given fixed term}
is decidable for right-shallow TRS and rewriting modulo permutative theories.
This result is used to obtain subsequent results. %
\\ (2)
In Section~\ref{sec-innermost},
we consider innermost rewriting modulo permutative
theories and show that termination is decidable for shallow TRS.
\\ (3)
In Section~\ref{sec-right-flat-linear}, we show that 
termination (and innermost termination as well) is decidable
for rewriting modulo permutative theories using TRS whose
right-hand side terms are both shallow and linear.
There is no restriction on the left-hand
side terms. Thus, right-ground TRS and shallow
right-linear TRS are both contained in our class.
\\ (4)
In Section~\ref{sec-hardness}, we prove that termination,
as well as innermost termination, is PSPACE-hard for
flat (and hence shallow) right-linear TRS.
\\ (5)
In Section~\ref{sec-undec},
we show undecidability of termination for
flat TRS and plain rewriting, and
undecidability of termination for right-shallow TRS
and innermost rewriting.

Uchiyama, Sakai and Sakabe~\cite{Sakai} have recently also 
generalized the results of our conference 
paper~\cite{DBLP:conf/rta/GodoyHT07} by replacing
syntactic restrictions on the rewrite rules by 
syntactic restrictions on the {\em{dependency pairs}}.
Specifically, termination and innermost termination were
shown to be decidable for TRS whose {dependency
pairs} are right-linear and right-shallow; and
innermost  termination was shown to be decidable for
TRS whose dependency pairs are shallow. 

\section{Preliminaries}\label{sec-preliminaries}

\noindent We use standard notation from the
term rewriting literature~\cite{Allthat}.
A signature $\Sigma$ is a (finite) set of function
symbols with arity, which is partitioned as $\cup_i\Sigma_i$
such that $f\in\Sigma_m$ if the arity of $f$ is $m$.
Symbols in $\Sigma_0$,
called {\em constants},
are denoted by $a,b,c,d,e$, with possible subscripts.
The elements of a set $\X$ of 
variable symbols are denoted by $x,y,z$ with possible subscripts. 
The set
$\Tau(\Sigma,\X)$
of {\em terms} over $\Sigma$ and $\X$,
is the smallest set containing $\X$ and
such that $f(t_1,\dots,t_m)$ is in $\Tau(\Sigma,\X)$
whenever $f \in \Sigma_m$,
and $t_1,\dots,t_m \in \Tau(\Sigma,\X)$. A
{\it position\/} is a sequence of positive integers.
The set of positions of a term $t$, denoted ${\tt Pos}(t)$,
is defined recursively as follows. If $t$ is a variable
then ${\tt Pos}(t)$ is $\{\lambda\}$, where $\lambda$
represents the empty sequence. If $t$ is of the
form $f(t_1,\ldots,t_m)$, then ${\tt Pos}(t)$ is
$\{\lambda\}\cup\{1.p|p\in{\tt Pos}(t_1)\}\cup\ldots\cup
\{m.p|p\in{\tt Pos}(t_m)\}$.
If $p$ is a
position and $t$ is a term, then by $t|_p$ we denote the {\em subterm
of $t$ at position $p$\/}: we have $t|_\lambda = t$ (where $\lambda$
denotes the empty sequence) and $f(t_1,\ldots,t_m)|_{i.p} = t_i|_p$ if
$1\leq i \leq m$ (and is undefined if $i>m$).
By $|p|$ we denote the length of a position $p$.
We also write $t[s]_p$
to denote the term obtained by replacing in $t$ the subterm at
position $p$ by the term $s$.
More formally, $t[s]_\lambda$ is $s$, and
$f(t_1,\ldots,t_{i-1},t_i,t_{i+1},\ldots,t_m)[s]_{i.p}$ is
$f(t_1,\ldots,t_{i-1},t_i[s]_{p},t_{i+1},\ldots,t_m)$.
For example, if $t$ is
$f(a,g(b,h(c)),d)$, then $t|_{2.2.1} = c$, and $t[d]_{2.2}=
f(a,g(b,d),d)$.
Note that $s=s[s|_p]_{p}$, and that the equality $s=s[u]_p$
implies $s|_p=u$.
The set of all subterms of a term $s$ is
denoted by $\subterms{s}$.
The symbol occurring at the root of a term $t$
is denoted as ${\tt root}(t)$.
We write $p_1 > p_2$ (equivalently, $p_2 < p_1$)
and say $p_1$ is below $p_2$
(equivalently, $p_2$ is above $p_1$) if $p_2$ is a proper
prefix of $p_1$, that is, $p_1 = p_2.p_2'$ for some non-empty $p_2'$.
In this case, by $p_1-p_2$ we denote $p_2'$.
By $p_1\leq p_2$ we denote that either $p_1<p_2$ or $p_1=p_2$ hold.
Positions $p$ and $q$ are {\em parallel}, denoted $p\parallel q$,
if $p\not\geq q$ and
$q\not\geq p$ hold.
By $\Vars(t)$ we denote the set of all variables occurring in a term $t$.
The {\tt height} of a term $s$ is $0$ if $s$ is a variable or a constant,
and $1+{\tt Max}_i({\tt height}(s_i))$ if $s=f(s_1,\ldots,s_m)$.
The {\tt depth} of an occurrence at position $p$ of a term $t$ in
a term $s=s[t]_p$ is $|p|$.
Sometimes we will denote a term $f(t)$ by
the simplified form $ft$ when the arity of $f$ is $1$, and $t[s]_p$ by
$t[s]$ when $p$ is clear from the context or not important.

A {\it substitution\/} $\sigma$ is a mapping from variables to terms.
It can be homomorphically extended to a function from
terms to terms:
$\sigma(t)$ denotes the result of
simultaneously replacing in $t$ every $x\in {\tt Dom}(\sigma)$ by $\sigma(x)$.
For example, if $\sigma$ is
$\{ x\mapsto f(b,y), y \mapsto a \}$, then
$\sigma(g(x,y))$ is $g(f(b,y),a)$.

A \emph{rewrite rule} over $\Sigma$
is a pair of terms $(l,r)$ of $\Tau(\Sigma,\X)$, denoted
by $l \to r$, with left-hand side $l$ and right-hand side $r$.
We make the usual assumptions for the rules, i.e.\ $l$ is not a variable,
and all variables occurring in the term $r$ also occur in the term $l$.
A \emph{term rewrite system} (TRS) $R$ over $\Sigma$ is a finite set of
rewrite rules over $\Sigma$. We often assume $\Sigma$ as implicit
when talking about a TRS $R$.
We say that $s$ rewrites to $t$ in one step at position $p$ (by $R$),
denoted by $s\to_{R,p} t$, if
$s|_p = \sigma(l)$ and $t = s[\sigma(r)]_p$, for some $l\to r\in R$
and substitution $\sigma$.
We also denote such a rewrite step by
$\rightarrow_{l\rightarrow r,p,\sigma}$
if we make explicit the used rule $l\rightarrow r$
and substitution $\sigma$.
If $p = \lambda$, then the rewrite step $\rightarrow_{R,p}$
is said to be applied at the root.
Otherwise, it is denoted by $s\rightarrow_{R,>\lambda} t$.

If $\to$ is a binary relation on a set $S$, then
$\leftrightarrow$ is its symmetric closure,
$\to^+$ is its transitive
closure, $\leftarrow$ is its inverse, and
$\to^*$ is its reflexive-transitive closure.

A \emph{(rewrite) derivation} (from $s$) is a sequence of
rewrite steps (starting from $s$), that is, a sequence
$s \tor s_1 \tor s_2 \tor \ldots$. With $s\rightarrow_R^* t$ we
denote that $t$ is $R$-reachable from $s$, or a concrete derivation
from $s$ to $t$, depending on the context.
A term $t$ is context-reachable from $s$ with $R$ (with a non-empty
context) if
there exists a derivation of the form $s\to_R^*u$ where
$t$ is a (proper) subterm of $u$.
The length of a derivation $s\to_R^*t$, denoted
$|s\to_R^*t|$, is its number of rewrite steps.
We denote this derivation as $s\to_R^0t$, $s\to_R^1t$
and $s\to_R^{0,1}t$ when this number is $0$, $1$, and
$0$ or $1$, respectively.
A TRS $R$ is {\em terminating from $s$}
if there are no {\em $R$-derivations},
$s \rightarrow_R s_1 \rightarrow_R \cdots$ with infinitely
many rewrite steps.
If $R$ is terminating from every term, then $R$ is 
said to be {\em terminating}.
A term $s$ is $R$-\emph{irreducible} (or, in {\em $R$-normal form}) 
if there is no term 
$t$ such that $s\rightarrow_R t$.
When there is a unique normal form reachable from a given term
$t$ using $R$, it is denoted by ${\tt NF}_R(t)$.
When $R$ is singleton, say $R = \{l\to r\}$, then
${\tt NF}_{R}(t)$ will also be written as ${\tt NF}_{l\to r}(t)$.

A term $t$ is called {\em ground} if $t$ contains no variables.
It is called {\em shallow} if all variable positions in $t$ are
at depth $0$ or $1$.
It is {\em flat} if its height is at most $1$.
It is {\em linear} if every variable occurs at most once.

A rule $l\to r$ is called {\em ground} ({\em flat}, {\em shallow},
{\em linear})
if both $l$ and $r$ are. A rule $l\to r$ is called
{\em left-ground} ({\em left-flat}, {\em left-shallow}, {\em
left-linear}) if $l$ is.
A rule $l\to r$ is called
{\em right-ground} ({\em right-flat}, {\em right-shallow},
{\em right-linear}) if $r$ is.
A rule $l\to r$ is called {\em collapsing} if $r$ is a variable.

A TRS $R$ is called (left-,right-){\em ground} ({\em flat}, {\em shallow},
{\em linear}) if all its rules are. A TRS $R$ is called {\em collapsing}
if it contains a collapsing rule.

A rewrite step
$s \rightarrow_{R, p} t$
is an {\em innermost rewrite step} if
$s|_{p'}$ is $R$-irreducible, for all $p'>p$.
The concepts of reachability and termination
can be naturally defined for innermost
rewriting.

A set $E$ of pairs of terms is a set of equations if
whenever a pair, again written as $l\to r$, belongs to $E$, 
the pair $r\to l$ also belongs to $E$.
Given a TRS $R$ and a set of equations $E$,
a term $s$ rewrites into a term $t$ with $R$ modulo $E$ in one step,
denoted $s\to_{R/E}t$, if $s\to_E^*\to_R\to_E^*t$ holds.
Note that $s\to_{R/E}^+t$ is equivalent to existence
of a derivation of the form $s\to_{R\cup E}^*t$ with
at least one rewrite step with $R$.
A {\em permutative rule} is a linear flat rewrite rule $l\to r$
satisfying ${\tt height}(l)={\tt height}(r)$ and $\Vars(l)=\Vars(r)$.
When $E$ contains just permutative rules we say that $E$
is a {\em permutative theory}.
In the rest of the paper
we will always assume that $E$ is a permutative theory
defined over the same signature as $R$.

The notion of innermost rewriting is extended to rewriting
modulo in the following natural way. A term $s$ is a normal
form with respect to $R/E$ if no $R/E$ rewrite step can
be applied on $s$. A term $s$ {\em{innermost rewrites}} to $t$
with $R/E$ if there exist terms $u,v$ and a position $p$ satisfying
$s\to_E^*u\to_{R,p}v\to_E^*t$ and such that any proper subterm
of $u|_p$ is a normal form with respect to $R/E$.

The notion of termination for $R/E$ is naturally defined
as the non-existence of a $R/E$ derivation with infinitely
many rewrite steps.
Note that this is equivalent to the non-existence of
a derivation with $R\cup E$ where infinitely many
of the involved steps use $R$.

\section{Flattening and Other Simplifying Assumptions}
\label{sec-flatten}

\noindent In this section we present some standard transformations
on the signature and TRS~\cite{GodoyTiwari05:CADE,DBLP:conf/rta/GodoyHT07}, 
and argue that they preserve
termination and innermost termination modulo permutative theories.
In particular, we show that we can assume without any loss of generality
that \\
(A1) the signature contains exactly one function symbol with nonzero arity
\\
(A2) all shallow terms are in fact flat.
\\
Readers who believe these claims can jump to the next section.

The discussion is
written for general termination, but it is also valid
when we interpret termination as innermost termination.
To this end, in the innermost case we assume that
for a given TRS, all the rules $l\rightarrow r$
such that $l$ has a proper subterm that is not a normal
form have been removed. Note that these rules can not
be used in an innermost derivation.
Thus, when considering innermost rewriting, we assume that
\\
(A3) if $l\rightarrow r$ is a rule in $R$, then all
proper subterms of $l$ are in normal form.

We will always assume that all terms
are constructed over a given fixed signature
$\Sigma$ that contains several constants and
only one non-constant function symbol $f$.
If this was not the case,
we can define a transformation $T$ from
terms over $\Sigma$ into terms over a new signature $\Sigma'$
as follows. Let $m$ be the
maximum arity of a symbol in $\Sigma$ plus $1$. We choose
a new function symbol $f$ with arity $m$ and define the new
signature $\Sigma'=\Sigma_0'\cup\Sigma_m'$ as
$\Sigma_0'=\Sigma$ and $\Sigma_m'=\{f\}$.
Note that all symbols of $\Sigma$ appear also in $\Sigma'$
but with arity $0$.
Now, we recursively define
$T:\Tau(\Sigma,\X)\rightarrow \Tau(\Sigma',\X)$
as $T(c)=c$ and $T(x)=x$ for constants $c\in\Sigma_0$ and variables $x\in\X$,
and $T(g(t_1,\ldots,t_k))=f(T(t_1),\ldots,T(t_k),g,\ldots,g)$
for terms headed with $g\in\Sigma-\Sigma_0$.
We denote $\{T(l)\rightarrow T(r)|l\rightarrow r\in R\}$ as $T(R)$
for a given TRS $R$.
Note that the size of $T(R)$ is at most $m$ times the
size of $R$, and hence, this transformation can be
easily performed in polynomial time.
Note also that $R$ is a TRS over $\Sigma$, and that $T(R)$
is a TRS over $\Sigma'$. As we mentioned in the preliminaries,
we will not explicitly state the signature of each TRS.

\begin{lemma}\label{lemma-simplifying1}
Let $R$ be a TRS. Let $E$ be a permutative theory.
Then, $R/E$ is (innermost) terminating if and only if $T(R)/T(E)$
is (innermost) terminating.
\end{lemma}

\proof
It is straightforward to see that, for any terms $s,t$
of $\Tau(\Sigma,\X)$,
$s\to_{R}t\Leftrightarrow T(s)\to_{T(R)}T(t)$
and $s\to_{E}t\Leftrightarrow T(s)\to_{T(E)}T(t)$ hold.
Thus, non-termination of $R/E$ trivially implies non-termination
of $T(R)/T(E)$.

For the left-to-right direction,
we define the 
transformation $T'$ as the following extension of the inverse of $T$
on the image. 
Since $T$ is not surjective, we will use two new function symbols,
$\$$ and $\#$ of arity $m$ and $0$ respectively, for defining
$T'(t)$ when $t$ is not in the image of $T$.
We define
$T':\Tau(\Sigma',\X)\rightarrow \Tau(\Sigma\cup\{\$,\#\},\X)$
as follows:
\begin{eqnarray*}
T'(c) = c, T'(x)=x & & \mbox{for constants $c\in\Sigma_0$ and variables $x\in\X$}
\\
T'(g)=\#  & & \mbox{for function symbols $g\in\Sigma-\Sigma_0$}
\\
T'(f(t_1,\ldots,t_k,g,\ldots,g)) = g(T'(t_1),\ldots,T'(t_k)) & & \mbox{for $g\in\Sigma_k$}
\\
T'(f(t_1,\ldots,t_m))=\$(T'(t_1),\ldots,T'(t_m)) & & \mbox{in all other cases}
\end{eqnarray*}

It is easy to see that any rewrite step
$s\to_{T(R)}t$ can be transformed into
a rewrite step $T'(s)\to_{T'(T(R))}T'(t)$,
and any rewrite step
$s\to_{T(E)}t$ can be transformed into
a rewrite step $T'(s)\to_{T'(T(E))}T'(t)$.
Thus, non-termination of $T(R)/T(E)$ implies non-termination
of $T'(T(R))/T'(T(E))$ for the signature
$\Sigma\cup\{\$,\#\}$. Note that $T'(T(R))$ and $T'(T(E))$
are, in fact, $R$ and $E$, respectively. Thus, we conclude that
$R/E$ is non-terminating over the signature
$\Sigma\cup\{\$,\#\}$. But, note that
non-termination (and non-termination modulo) of a TRS
does not depend on symbols in the signature that
do not occur in the rules. Hence, $R/E$ is non-terminating
over the original signature, and we are done.
\qed

In the case where $R$ is left-shallow, we will also assume
that $R$ is, indeed, left-flat. If this was not the case,
we proceed by applying several times the following
transformation step a), until $R$ is left-flat.

\medskip

\noindent
{\bf step a)} If there is a non-constant ground term $u$ that is
a proper subterm of a left-hand side of a rule in $R$, then create
a new constant $c$, replace all occurrences of $u$ in
the left-hand sides of the rules of $R$ by $c$, and add the
rule $u\rightarrow c$ to $R$. Formally,
the new TRS $R'$ is
$\{u\to c\}\cup\{{\tt NF}_{u\to c}(l)\to r|\;(l\to r)\in R\}$.
Note that, as a consequence of Assumption~(A3),
when considering innermost rewriting, $u$ is
necessarily a normal form.

\medskip

We will also assume that all rules in $R$ are right-flat.
If this was not the case, as before
we proceed by applying several times the following
transformation step b), until the obtained TRS is right-flat.

\medskip

\noindent
{\bf step b)} If there is a non-constant ground term $u$ that is a proper
subterm of a right-hand side of a rule in $R$, then create
a new constant $c$, replace all occurrences of $u$ in
the right-hand sides of the rules of $R$ by $c$, and add the
rule $c\rightarrow u$ to $R$. Formally,
the new TRS $R'$ is
$\{c\to u\}\cup\{l\to{\tt NF}_{u\to c}(r)|\;(l\to r)\in R\}$.

\medskip

Every step (a or b) decreases the total sum of the number of positions at
depth more than one in all the left-hand and right-hand sides of $R$.
Moreover, also at every step, the total size of the TRS
increases by at most the size of two constants.
Hence, this process terminates in linear time and the size of
the resulting flat TRS is within a constant factor of the size of
the original shallow TRS.

\begin{lemma}\label{lemma-simplifying2}
Let $R$ be a TRS. Let $E$ be a permutative theory.
Let $R'$ be obtained from $R$ by applying step a).
Then, $R/E$ is (innermost) terminating if and only if
$R'/E$ is (innermost) terminating.
\end{lemma}
\proof
For the right-to-left direction, we first observe
that each rewrite step $s\to_Rt$ can be transformed
into a derivation of the form $s\to_{R'}^+t$,
since the application of a rewrite rule $l\to r$
can be simulated by several applications of $u\to c$
and one application of ${\tt NF}_{u\to c}(l)\to r$.
Thus, any derivation of $R\cup E$ with infinitely many
rewrite steps of $R$ and starting from a certain term $s$
can be transformed into a derivation of $R'\cup E$
with infinitely many rewrite steps of $R'$ and starting from $s$.
In the case of innermost rewriting,
we have that $u$ is a normal form and hence, 
the transformed derivation is also innermost.

For the left-to-right direction, we first observe the
following two facts:
\begin{enumerate}[$\bullet$]
\item The existence of a rewrite step $s\to_{u\to c}t$ implies
${\tt NF}_{c\to u}(s)={\tt NF}_{c\to u}(t)$.
\item For each rule $l\to r$ of $R$, 
if $l'={\tt NF}_{u\to c}(l)$, then, for each rewrite step
$s\to_{l'\to r}t$, it holds that
${\tt NF}_{c\to u}(s)\to_{l\to r}{\tt NF}_{c\to u}(t)$.
\end{enumerate}
From the above facts, it follows that any rewrite step
$s\to_{R'}t$ can be transformed into a derivation 
${\tt NF}_{c\to u}(s)\to_R^* {\tt NF}_{c\to u}(t)$
with $0$ or $1$ steps. Note that this is not enough to
argue that a derivation of $R'\cup E$ with infinitely
many steps with $R'$, and starting
from a term $s$, can be transformed into a derivation
of $R\cup E$ with infinitely many steps with $R$ starting
from ${\tt NF}_{c\to u}(s)$. This is because
rewrite steps with $u\to c$ are, in fact, removed. 
However, it suffices to additionally argue that
a derivation of $\{u\to c\}\cup E$
with infinitely many steps of $\{u\to c\}$ cannot exist.
This is a consequence of the fact
that rules of $E$ preserve the size, and $u\to c$ decreases
the size.
Finally, in case of innermost rewriting, using the facts
that $u$ is a normal form, $s\to_{R'} t$ is an innermost
step, and $l' = {\tt NF}_{u\to c}(l)$, we infer that 
the transformed derivation is an innermost derivation.
\qed

The preservation of termination for the case of step b)
is proved analogously.

\begin{lemma}\label{lemma-simplifying3}
Let $R$ be a TRS. Let $E$ be a permutative theory.
Let $R'$ be obtained from $R$ by applying step b).
Then, $R/E$ is (innermost) terminating if and only if $R'/E$
is (innermost) terminating.\qed
\end{lemma}

\section{Right Flat TRS}\label{sec-right-flat}

\noindent In this section, we will show that, given a 
right-flat TRS $R$ and a term $s$,
it is decidable if $R$ is terminating from $s$.
In particular, this implies that
non-termination is semi-decidable for
right-flat TRS.
We will show that termination is undecidable
for right-flat TRS in Section~\ref{sec-undec}.

The proofs of this section are
written for general termination, but they are also valid
when we interpret termination as innermost termination,
reachability as innermost reachability, and so on.

An important property of a right-flat TRS $R$ is that
whenever $s \rightarrow_R^* t$ holds, then every subterm of $t$ is
reachable from either a constant or some subterm of $s$.
This result, stated as Lemma~\ref{lemma-reach1}, is 
used extensively later.  It is proved 
by inductively 
marking each position of a term (in the above derivation) by
a term from $\subterms{s}\cup\Sigma_0$. The idea of
the marking $t_p$ at each position $p$ of $t$ is that
it satisfies $t_p\rightarrow_R^* t|_p$, and moreover,
for a position $p'>p$ the corresponding marking
$t_{p'}$ of $t$ at $p'$ is context-reachable
from $t_p$.

\begin{defi}
Let $R$ be a right-flat TRS.
Let $s=s_1\to_R\ldots\to_Rs_n$ be a derivation with $R$.
A {\em Marking} of this derivation is a sequence
$M_1,\ldots,M_n$ of $n$ functions
$M_i:\Pos(s_i)\mapsto \subterms{s}\cup\Sigma_0$
defined inductively as follows:
\begin{enumerate}[$\bullet$]
\item For each $p$ in ${\tt Pos}(s)$, we define $M_1(p)=s|_p$.
\item For $1\leq i< n$ we assume that $M_i$ 
is defined. Let $s_i\to_{l\to r,\bar{p}}s_{i+1}$ be the $i$'th
rewrite step of the derivation above more explicitly written. Then,
we define $M_{i+1}$ as follows:
\begin{enumerate}[(i)]
\item For each $p$ in ${\tt Pos}(s_{i+1})$
satisfying $p\not> \bar{p}$, we define $M_{i+1}(p) := M_i(p)$.
\item For each $p$ in ${\tt Pos}(s_{i+1})$
satisfying $p=\bar{p}.p_0$, $|p_0|=1$, and $r|_{p_0}$ is a constant, 
we define $M_{i+1}(p) := r|_{p_0}$.
\item For each $p$ in ${\tt Pos}(s_{i+1})$ 
satisfying $p = \bar{p}.p_0.p_1$, $|p_0|=1$, $|p_1|\geq 0$, and $r|_{p_0}$ is a variable,
we define $M_{i+1}(p) := M_i(\bar{p}.q_0.p_1)$, where
$q_0$ is any position in $l$ such that $l|_{q_0} = r|_{p_0}$.
\end{enumerate}
\end{enumerate}
\end{defi}

\noindent Recall that we are assuming that every variable on the right-hand
side also appears on the left-hand side; if not, then the TRS
is trivially non-terminating.
The following example illustrates the definition of marking and
also shows that markings need not be unique.

\begin{example}[Marking]
\label{example-marking}
Let $R := \{a\to f(c), b\to f(c), g(x,x)\to f(x)\}$ and consider
the derivation
$s_1 := g(a,b) \to s_2 := g(f(c),b) \to s_3 := g(f(c),f(c)) \to s_4 := f(f(c))$.
A marking for this derivation is given by:
$M_1(p) = s_1|_p$ for all $p\in\Pos(s_1)$,
$M_2(p) = M_1(p)$ for all $p\in\Pos(s_2)-\{11\}$,
$M_2(11) = c$,
$M_3(p) = M_2(p)$ for all $p\in\Pos(s_3)-\{21\}$,
$M_3(21) = c$, 
$M_4(\lambda) = s_1$,
$M_4(1) = a$, and
$M_4(11) = c$.
Note that if we redefined $M_4(1)$ so that
$M_4(1) = b$, then the resulting functions would still be a marking.
Hence, there can be multiple markings for the same derivation.
\end{example}

Now we will state and prove some useful properties about markings.
Henceforth, let us fix $R$ to be a right-flat TRS,
$s=s_1 \to_R\ldots\to_Rs_n$ to be a (innermost) derivation 
and
$M_1,\ldots,M_n$ to be a marking of this derivation.
The properties below will capture the intuition
that, if $M_i(p) = t$, then the term $s_i|_{p}$ 
is reachable from the term $t$.

\begin{lemma}\label{lemma-reach0}
$M_n(\lambda) = s$.
Moreover, if $s$ is not a constant, then,
for each $p$ in $\Pos(s_n)-\{\lambda\}$ we have $M_n(p)\neq s$.
\end{lemma}

\proof
The claim is proved %
by induction on $n$.
For $n=1$, by definition of marking, we have
$M_1(\lambda)=s_1|_\lambda=s$. Moreover, if $s$ is not a constant,
for each $p$ in ${\tt Pos}(s_1)-\{\lambda\}$ we have
$M_1(p)=s_1|_p=s|_p\not=s$.

For the induction step,
suppose $s_n \rightarrow_{l\rightarrow r,\bar{p}} s_{n+1}$.
By induction hypothesis, we know that $M_n(\lambda) = s$ and
whenever $s$ is not a constant then, for each $p\in{\tt Pos}(s_n)-\{\lambda\}$,
$M_n(p)\not=s$ holds.
The fact that $M_{n+1}(\lambda)=s$ follows from the fact
that $M_n(\lambda)=s$ and $M_{n+1}(\lambda)=M_n(\lambda)$
holds, since Case (i) of the definition of marking applies for
$p=\lambda$.
Under the assumption that $s$ is not a constant, we note
that Case (ii) defines $M_{n+1}(p)$ as a constant, and cases (i)
and (iii) define $M_{n+1}(p)$ for $p$ in ${\tt Pos}(s_{n+1})-\{\lambda\}$ as
$M_n(p')$ for some $p'$ in ${\tt Pos}(s_n)-\{\lambda\}$.
Thus, from the assumption
that $M_n(p')\not= s$, it follows that $M_{n+1}(p)\not= s$.\qed

A second property of markings is that
$s_n|_p$ is always reachable from $M_n(p)$.
\begin{lemma}\label{lemma-reach1}
For each $p$ in $\Pos(s_n)$,
$s_n|_p$ is (innermost) reachable from $M_n(p)$ .
\end{lemma}
\proof
The claim is proved %
by induction on $n$.
For $n=1$, by definition of marking, we have $M_1(p)=s_1|_p$ for
each $p$ in ${\tt Pos}(s_1)$. Thus,
$M_1(p)\to_R^*s_1|_p$ in $0$ steps follows trivially.

For the induction step,
suppose $s_n \rightarrow_{l\rightarrow r,\bar{p}} s_{n+1}$ is the
$n$-th (innermost) rewrite step.
By induction hypothesis, $M_n(p)\to_R^*s_n|_{p}$ holds for each $p\in{\tt Pos}(s_n)$.
Consider a fixed $p\in\Pos(s_{n+1})$. We prove
$M_{n+1}(p)\to_R^*s_{n+1}|_{p}$ as follows:

\begin{enumerate}[$\bullet$]
\item If $p\leq \bar{p}$, then we have
$M_{n+1}(p) = M_{n}(p)\rightarrow_R^* s_n|_{p}\rightarrow_{l\rightarrow
r,\bar{p}-p} s_{n+1}|_{p}$. 
Note that, since $s_n$ (innermost) rewrites to $s_{n+1}$, it follows that
$s_{n}|_{p}$ (innermost) rewrites to $s_{n+1}|_{p}$.

\item If $p = \bar{p}.p_0$ and $|p_0|=1$ hold, and $r|_{p_0}$ is a constant, then,
by definition of marking we have $M_{n+1}(p)$ is $r|_{p_0}$,
from which $M_{n+1}(p)\to_R^*r|_{p_0}=s_{n+1}|_{p}$
in $0$ steps follows trivially.

\item If $p = \bar{p}.p_0.p_1$ and $|p_0.p_1|\geq 1$ hold,
and $r|_{p_0}$ is a variable,
then, for some $q_0$,
$M_{n+1}(p) = M_{n}(\bar{p}.q_0.p_1)$ and
$s_{n+1}|_{p} = s_n|_{\bar{p}.q_0.p_1}$ hold, and
$M_{n}(\bar{p}.q_0.p_1)\to_R^*s_n|_{\bar{p}.q_0.p_1}$ holds
by induction hypothesis. Thus,
$M_{n+1}(p)\to_R^*s_{n+1}|_{p}$ follows.

\item If $p\parallel \bar{p}$, then the claim holds by induction
hypothesis again as
$s_{n+1}|_{p}=s_n|_{p}$ and $M_{n+1}(p)=M_{n}(p)$ hold.
\end{enumerate}
Thus, for each position $p\in\Pos(s_{n+1})$, we proved that
$s_{n+1}|_p$ is (innermost) reachable from $M_{n+1}(p)$ .
\qed

\begin{cor}\label{corollary-reach1}
If $s$ is a constant, then all subterms of $s_n$
are (innermost) reachable from a constant.\qed
\end{cor}

Another property of markings is that 
$M_n(p)$ is context-reachable from $M_n(p')$ 
for all $p' \leq p$.
\begin{lemma}\label{lemma-creach1}
For %
each $p,p'\in\Pos(s_n)$ satisfying $p' < p$,
$M_n(p)$ is (innermost) context-reachable from $M_n(p')$.
Moreover, if 
$M_n(p)$ and $M_n(p')$ are both constants,
then $M_n(p)$ is (innermost) context-reachable from $M_n(p')$
with a non-empty context.
\end{lemma}
\proof
The claim is proved %
by induction on $n$.
For $n=1$, by definition of marking we have $M_1(p)=s_1|_p$ for
each $p$ in ${\tt Pos}(s_1)$. Since, for each
$p,p'\in\Pos(s_1)$ satisfying $p' < p$, 
$s_1|_{p'}=s_1|_{p'}[s_1|_p]_{p-p'}$ holds, then we also have
$M_1(p')=M_1(p')[M_1(p)]_{p-p'}$. Thus, the statement
trivially follows for the base case.

For the induction step,
suppose $s_n \rightarrow_{l\rightarrow r,\bar{p}} s_{n+1}$ is
the $n$'th (innermost) rewrite step. %
Consider two fixed positions
$p,p'\in\Pos(s_{n+1})$ satisfying $p' < p$.
We distinguish the following cases.
\begin{enumerate}[$\bullet$]
\item If $p\leq \bar{p}$ or $p\parallel \bar{p}$, then we have
$M_{n+1}(p) = M_n(p)$ and $M_{n+1}(p') = M_n(p')$.
Thus, the statement follows by induction hypothesis.

\item If $p = \bar{p}.p_0$, $|p_0|=1$, and $r|_{p_0}$ is a constant, then
$M_{n+1}(p) = r|_{p_0}$ holds.
By Lemma~\ref{lemma-reach1},
$s_{n+1}|_{p'}$ is reachable from $M_{n+1}(p')$.
Note that $M_{n+1}(p)$ is a proper subterm of $s_{n+1}|_{p'}$.
Hence,
$M_{n+1}(p)$ is context reachable from $M_{n+1}(p')$
with a non-empty context (independently of whether
$M_{n+1}(p')$ is a constant or not).

\item If $p = \bar{p}.p_0.p_1$, $|p_0.p_1|\geq 1$ and $r|_{p_0}$ is a variable,
then, for some $q_0$, $M_{n+1}(p) = M_{n}(\bar{p}.q_0.p_1)$.
We distinguish two cases. 
(a) If $p' \leq \bar{p}$, then 
 $M_{n+1}(p') = M_{n}(p')$ and, by induction hypothesis,
$M_n(\bar{p}.q_0.p_1)$ is context reachable from $M_n(p')$
(with a non-empty context if both $M_n(p')$ and
$M_n(\bar{p}.q_0.p_1)$ are constants),
which is the same as saying that
$M_{n+1}(p)$ is context reachable from $M_{n+1}(p')$
(with a non-empty context if both $M_{n+1}(p')$ and
$M_{n+1}(p)$ are constants).
(b) If $p' > \bar{p}$ holds, then
$M_{n+1}(p') = M_n(\bar{p}.q_0.p_1')$ holds for some $p_1' < p_1$
and, by induction hypothesis,
$M_n(\bar{p}.q_0.p_1)$ is context reachable from
$M_n(\bar{p}.q_0.p_1')$
(with a non-empty context if both $M_n(\bar{p}.q_0.p_1')$ and
$M_n(\bar{p}.q_0.p_1)$ are constants).
This is the same as saying that
$M_{n+1}(p)$ is context reachable from $M_{n+1}(p')$
(with a non-empty context if both $M_{n+1}(p')$ and
$M_{n+1}(p)$ are constants).
\end{enumerate}
Thus, in all cases, the claim follows. \qed

We illustrate Lemma~\ref{lemma-creach1} by an example below.
\begin{example}[Lemma~\ref{lemma-creach1}]
Consider again the derivation and marking defined in
Example~\ref{example-marking}.
We note that, on the term $s_4 := f(f(c))$,
we had the marking $M_4$ defined so that
$M_4(1) = a$ and $M_4(11) = c$.
By Lemma~\ref{lemma-creach1}, $c$ should be context-reachable
from $a$, and indeed we have $a \to f(c)$.
\end{example}

Finally, another observation about a marking is 
that positions below ${\tt height}(s)$ are always
marked by constants.
\begin{lemma}\label{lemma-creach2}
For each $p\in\Pos(s_n)$ such that $|p| >{\tt height}(s)$,
$M_n(p)$ is a constant.
\end{lemma}
\proof
The claim is proved %
by induction on $n$.
For $n=1$, note that $s_1=s$ holds and hence all $p\in\Pos(s_1)$
satisfy $|p|\leq{\tt height}(s)$. Thus, the claim is vacuously true.

For the induction step,
suppose $s_n \rightarrow_{l\rightarrow r,\bar{p}} s_{n+1}$ is
the $n$'th (innermost) rewrite step. %
Consider any position $p\in\Pos(s_{n+1})$ satisfying
$|p| > {\tt height}(s)$.
\begin{enumerate}[$\bullet$]
\item If $p\parallel \bar{p}$ or $p\leq \bar{p}$,
then $M_{n+1}(p) = M_n(p)$ holds, and by induction hypothesis
$M_n(p)$ is a constant.
\item
If $p = \bar{p}.p_0$, $|p_0|=1$ and $r|_{p_0}$ is a constant, 
then $M_{n+1}(p) = r|_{p_0}$ holds, which is a constant.
\item
If $p = \bar{p}.p_0.p_1$, $|p_0.p_1|\geq 1$ and $r|_{p_0}$ is a variable, then
$M_{n+1}(p) = M_{n}(\bar{p}.q_0.p_1)$ for some $q_0$, and
$|\bar{p}.q_0.p_1| \geq |p|$ holds since left-hand sides of $R$
are not variables and $R$ is right-flat.
Hence, the induction hypothesis is applicable and
we can conclude that $M_n(\bar{p}.q_0.p_1)$, and therefore
$M_{n+1}(p)$, is a constant.
\end{enumerate}
Thus, for all $p$ s.t. $|p|>{\tt height}(s)$, $M_{n+1}(p)$ is a constant.
This completes the proof. \qed

An important consequence of Lemma~\ref{lemma-creach1}
and Lemma~\ref{lemma-creach2}
is that, if $R$ is terminating from $s$,
then the height of terms reachable from $s$ is
bounded by the height of $s$ plus the number of
constants in $\Sigma$.
\begin{cor}\label{corollary-bounded}
Let $R$ be a right-flat TRS.
Let $E$ be a permutative theory.
Let $s$ be a term
such that $R/E$ is (innermost) terminating from $s$.
Then for any term $t$ (innermost) reachable from $s$ with $R/E$,
we have ${\tt height}(t) \leq {\tt height}(s) + |\Sigma_0|$.
\end{cor}
\proof
We proceed by contradiction by assuming
$s\to_{R/E}^*t$ and ${\tt height}(t) > {\tt height}(s) + |\Sigma_0|$.
Recall that the derivation $s\to_{R/E}^*t$ can be seen
as a derivation $s\to_{R\cup E}^*t$.
Let $M_1,\ldots,M_n$ be a marking of this derivation
$s\rightarrow_{R\cup E}^* t$.
By Lemma~\ref{lemma-creach2}, each position in $t$ that
is deeper than ${\tt height}(s)$ is marked with a constant.
Since
${\tt height}(t) > {\tt height}(s) + |\Sigma_0|$ holds,
by pigeon-hole principle, there are two positions
$p,p'\in\Pos(t)$ such that
$p < p'$ and $M_n(p) = M_n(p')$ hold, and $M_n(p)$ is a constant, say $c$.
By Lemma~\ref{lemma-creach1}, it follows that
$c$ is context reachable from $c$ with a non-empty context.
Moreover, since $E$ is a permutative theory,
$c$ is context reachable from $c$ with a derivation using
at least one rewrite step with a rule of $R$.
Furthermore,
by Lemma~\ref{lemma-reach0} the position $\lambda$ of every term in
a derivation is marked with $s$.
Using Lemma~\ref{lemma-creach1} again,
we infer that $M_n(p)=c$, is also context reachable from $s$.
Thus, we can construct a derivation
$s \rightarrow_{R\cup E}^* C_1[c] \rightarrow_{R\cup E}^+
C_1[C_2[c]] \rightarrow_{R\cup E}^+ C_1[C_2[C_2[c]]]
\rightarrow_{R\cup E}^+ \ldots$ with infinitely many
steps with $R$.
Hence, there is a derivation starting from $s$ using $R/E$
with infinitely many rewrite steps, a contradiction. \qed

Using the above corollary, we can show that 
the existence of non-terminating derivations 
starting from a term is decidable for right-flat TRS. 
\begin{thm}\label{theorem-c}
Termination (innermost termination) of a right-flat TRS $R$
modulo a permutative theory $E$ from
a given term is decidable. Hence, non-termination 
(innermost non-termination) is
semi-decidable for right-flat TRS modulo permutative theories.
\end{thm}
\proof
Let $s$ be any term.
We enumerate all (innermost) derivations starting from $s$.
If we reach a term with height greater than
${\tt height}(s) + |\Sigma_0|$,
then by Corollary~\ref{corollary-bounded} we know
that $R/E$ is non-terminating from $s$.
Otherwise, we will get only {\em finitely} many
reachable terms.  If there is a derivation that 
cycles among these terms, then $R/E$ is non-terminating
from $s$. If not, then $R/E$ is terminating from $s$.  \qed

\noindent
{\em Remark:}
We can use an argument similar to the one used in the proof of 
Theorem~\ref{theorem-c} to prove that,
for any class ${\mathcal C}$ of TRS's that are effectively
regularity preserving,
termination of a TRS $R$ of ${\mathcal C}$
from a term $s$, where both $R$ and $s$ are given as input, 
is decidable.
While we do not use this observation here,
we nevertheless note that, using recent results on
regularity preserving TRSs~\cite{TakaiKajiSeki00:RTA},  
we immediately get very simple proofs of known decidability results, such as for
right-ground TRS~\cite{Dershowitz81:ICALP}: a right-ground
TRS is regularity preserving, and is non-terminating iff it is 
non-terminating from some right-hand side, which can
be checked for every right-hand side term using the above
observation.

\ignore{
We state below a related new (up to our knowledge) theorem that
uses a similar argument as the proof
of Theorem~\ref{theorem-c}.
\begin{thm}\label{theorem-unused}
Let ${\mathcal C}$ be a class of TRS's that are effectively
regularity preserving.
Termination of a TRS $R$ of ${\mathcal C}$
from a term $s$, where both $R$ and $s$ are given as input, is decidable.
\end{thm}
\proof
Note that $\{s\}$ is a regular tree language.
Since $R$ is effectively regularity preserving,
we can compute a tree automaton $A$ recognizing the
set of terms reachable from $\{s\}$.
The size of the language recognized by $A$ can be
checked to be infinite, in which case we know that there exists
a derivation starting from $s$ with infinitely many
rewrite steps. Otherwise, we have a finite
number of terms reachable from $s$. By producing all possible
derivations starting from $s$ we will either detect a cycle,
thus concluding non-termination, or will halt, thus concluding
termination.\qed

We shall not use Theorem~\ref{theorem-unused} in this paper.
However, we note here that, using recent results on
regularity preserving TRS~\cite{TakaiKajiSeki00:RTA},  
we immediately get very simple
proofs of known decidability results, such as for
right-ground TRS~\cite{Dershowitz81:ICALP}: a right-ground
TRS is regularity preserving, and is non-terminating iff it is 
non-terminating from some right-hand side, which can
be checked for every one using Theorem~\ref{theorem-unused}.
\endignore}

\section{Innermost Termination of Flat TRS's}\label{sec-innermost}

\noindent In this section, we show that innermost termination 
of flat TRS modulo permutative theories is decidable. In sharp contrast,
general termination is undecidable for 
flat TRS (Section~\ref{sec-undec}).

Let $R$ be a flat TRS, and let $E$ be a permutative theory.
We show decidability of innermost termination of
$R/E$ by showing that 
if $R/E$ is not innermost terminating, then
there will be an infinite $R/E$ derivation starting 
from a ground flat term.
Using Theorem~\ref{theorem-c}, we know that this latter 
check is decidable.

\begin{lemma}\label{lemma-flat-innermost}
Let $R$ be a flat TRS.
Let $E$ be a permutative theory.
Suppose that $R/E$ is not
innermost terminating. Then, there is an
innermost derivation starting from
a ground flat term with infinitely many innermost rewrite steps.
\end{lemma}
\proof
We assume that
there is no innermost derivation with infinitely many
innermost rewrite steps and starting from a constant, and we
show that there is one from a ground flat term with height $1$.

Since $R/E$ is not innermost terminating,
there exists an
innermost derivation
$t_0\rightarrow_{R\cup E} t_1\rightarrow_{R\cup E} \ldots$
with infinitely many innermost rewrite steps using $R$,
whose first step is at position $\lambda$. 
We first prove that for every $i$,
{\em every subterm at depth $1$ of $t_i$ is either
reachable from a constant, or a normal form.}
First note that no term
$t_i$ is a constant, by our initial assumption.
Moreover, since we use innermost rewriting,
all proper subterms of $t_0$ are normal forms. 
By Lemma~\ref{lemma-reach1}, all subterms at depth 1
of $t_i$ are innermost reachable from either constants or proper
subterms of $t_0$. Hence
they are innermost reachable from constants, or they are normal forms.

Now, we note that there exists at least one constant,
call it $c$, that is
a normal form.  If not, any ground term can be 
innermost rewritten to another ground term, and 
hence there will be innermost derivations starting from
constants with infinitely many innermost rewrite steps,
which contradicts our initial assumption.

We construct a new innermost derivation
$t_0'\rightarrow_{R\cup E,\lambda} t_1'\rightarrow_{R\cup E}\ldots$
by defining each $t_i'$ to be as $t_i$ but replacing every
subterm at depth $1$ that is not innermost reachable from any constant
by the constant $c$ chosen above. We need to show that 
the new derivation is ``correct'', that is,
there is an innermost rewrite step from $t_{i-1}'$ to $t_{i}'$.
Consider the corresponding innermost rewrite step 
$t_{i-1}\rightarrow_{l\rightarrow r,\bar{p}} t_i$.
\begin{enumerate}[$\bullet$]
\item If $\bar{p}$ is not $\lambda$, then $\bar{p}$ is
of the form $j.p$ for some $j$ in $\{1,\ldots,m\}$
and some position $p$. Since $t_{i-1}|_j$ is rewritten,
it is not a normal form. Thus it is innermost reachable from a constant,
and hence, $t'_{i-1}|_j$ and $t'_{i}|_j$ coincide with
$t_{i-1}|_j$ and $t_i|_j$, respectively.
Therefore, the same innermost rewrite step can be applied on $t_{i-1}'$
to produce $t_i'$.
\item If $\bar{p}$ is $\lambda$, then,
by our initial assumption, both $l$ and $r$ are not constants.
Moreover, $r$ cannot be a variable, since, 
otherwise, $t_i$ would be a normal form
since we have innermost rewriting (and the derivation would
be finite).
Hence, $l\rightarrow r$ is of the form
$f(\alpha_1,\ldots,\alpha_m)\rightarrow f(\beta_1,\ldots,\beta_m)$.
Recall that, since $R$ is flat, each $\alpha_i$ and each $\beta_i$
is either a constant or a variable.
If $\sigma$ is the substitution used in this innermost rewrite step,
then define $\sigma'$ to be as $\sigma$ except for the cases where 
$\sigma(x)$ is not innermost reachable from a constant, in which case we
define $\sigma'(x)=c$. With these definitions,
$t_{i-1}' \rightarrow_{l\rightarrow r,\sigma',\lambda} t_i'$
is an innermost rewrite step.
\end{enumerate}
The derivation $t_0'\rightarrow_{R\cup E}
t_1'\rightarrow_{R\cup E}\ldots$ is again innermost,
has infinitely many innermost rewrite steps with $R$, and the
initial term $t_0'$ satisfies that all its
subterms at depth $1$ are innermost reachable from constants.
Therefore, there exists a ground flat term $s$ with height $1$ such that
$s\rightarrow_{R\cup E}^* t_0'$ is an innermost derivation,
and hence, there exists
an innermost derivation with infinitely many innermost rewrite
steps starting from a ground flat term $s$
with height $1$.  \qed

\begin{thm}\label{theorem-innermost}
Innermost termination modulo permutative theories is decidable
for shallow TRS's.
\end{thm}
\proof
By Lemmas~\ref{lemma-simplifying1}, \ref{lemma-simplifying2}
and~\ref{lemma-simplifying3}
innermost termination of shallow TRS modulo permutative theories can
be reduced to the particular case where $R$ is flat
and where the signature contains just one non-constant function symbol.

Since there are only finitely many ground
flat terms, using Theorem~\ref{theorem-c}, we can check
if a given flat $R/E$ is not innermost terminating starting from
one of these terms. By Lemma~\ref{lemma-flat-innermost},
we will find a witness for non-termination this way
iff $R/E$ is not innermost terminating.
\qed

\section{Termination and Innermost Termination of Right-Flat Right-Linear TRS's}
\label{sec-right-flat-linear}

\noindent In this section, we show decidability of termination and
innermost termination for right-flat right-linear TRS.
Again, the proofs of this section are written for
general rewriting, but they remain valid for
innermost rewriting.

The proof of decidability of (innermost) termination for
right-flat right-linear TRS depends on two key
observations. The first one is Lemma~\ref{lemma-reach1},
which says that for any (innermost) derivation 
$s\rightarrow_R^* t$ using a right-flat TRS $R$,
every {\em proper} subterm of
$t$ is (innermost) reachable from either a constant or a 
{\em proper} subterm of $s$.
The second key lemma is stated by first
defining the following {\em measure} of a term t: 
$$
\|t\|:=|\{p\;\mid\;p\in{\tt Pos}(t)\;\wedge\;t|_p\mbox{ is not (innermost) reachable from a constant }\}|
$$
Note that $\|t\|$ depends on
whether we are dealing with general or innermost rewriting.

Let us fix $R$ to be a right-flat right-linear TRS and
$E$ to be a permutative theory.
The first lemma below uses right-linearity of $R$.

\begin{lemma}\label{decreasing}
If $s\to_R t$, then $\|s\|\geq\|t\|$.
Moreover, if $s[f(s_1,\ldots,s_m)]_{\bar{p}}$ rewrites to $t$
at position $\bar{p}$ with a rule $f(l_1,\ldots,l_m)\rightarrow r$,
and $\|s\|=\|t\|$, then, for every $i$ in $\{1\ldots m\}$,
if $s_i$ is not reachable from a constant, then
$l_i$ is a variable.
\end{lemma}
\proof
Let $s\rightarrow_{l\rightarrow r,\bar{p}} t$ be the rewrite
step of the lemma.
We prove the first statement by constructing
an injective map, from positions $p$ of ${\tt Pos}(t)$
such that $t|_{p}$ is not reachable from a constant, to
positions $p'$ of $s$ such that $s|_{p'}$ is not reachable
from a constant, as follows. If $p\parallel \bar{p}$ or $p\leq \bar{p}$,
then we let $p':=p$. If $p>\bar{p}$, then $p$ can be written
in the form $\bar{p}.p_0.p_1$ where $r|_{p_0}$ is a height $0$ term.
In fact, $r|_{p_0}$ cannot be a constant since otherwise
$t|_{p}$ would be a constant. Hence, $r|_{p_0}$ is a variable.
We choose a position $p_0'$ such that $l|_{p_0'}$ is the
same variable as $r|_{p_0}$ and define $p' :=\bar{p}.p_0'.p_1$. The injectivity
of the map follows by right-linearity of $R$. Hence,
$\|s\| \geq \|t\|$ holds.

For the second statement,
we assume $\|s\|=\|t\|$, that
$s$ is of the form $s[f(s_1,\ldots,s_m)]_{\bar{p}}$, and $l$ is
of the form $f(l_1,\ldots,l_m)$. If a certain $s_i$ is not
reachable from a constant, but $l_i$ is not a variable, then
$\bar{p}.i$ is not in the image of the previous mapping, and hence
$\|s\|>\|t\|$ holds, contradicting $\|s\|=\|t\|$.
Therefore, all such $l_i$'s are variables.  \qed

Note that since $E$ is linear and flat, Lemma~\ref{decreasing}
applies to rewrite steps with $E$ too.
In the next lemma, if $R/E$ is non-terminating,
we establish the existence of 
a non-terminating derivation with certain properties.

\begin{lemma}\label{lemma-main-aux}
If $R/E$ is (innermost) non-terminating  and
there is no (innermost) non-terminating derivation starting
from a constant, then there is an infinite derivation
$t_0\rightarrow_{R\cup E} t_1\rightarrow_{R\cup E} \ldots$
with infinitely many rewrites with $R$ and
with the following properties:
\begin{enumerate}[\em(a)]
\item there is no infinite derivation starting from a proper
subterm of $t_0$

\item there is no rewrite with a collapsing rule at position $\lambda$

\item there are infinitely many rewrites at position $\lambda$
\end{enumerate}
\end{lemma}
\proof
Since $R/E$ is non-terminating,
there exists a
derivation $t_0\rightarrow_{R\cup E} t_1\rightarrow_{R\cup E} \ldots$
with infinitely many rewrite steps with $R$.
We pick the derivation that has minimal height for $t_0$.
We claim this derivation has all the properties mentioned above.

It has Property~(a) due to our choice of the infinite derivation.
Next assume that 
$t_{i-1}\rightarrow_R t_i$ is the first 
application of a collapsing rule at $\lambda$ in the derivatin.
Then,
by Lemma~\ref{lemma-reach0} and
Lemma~\ref{lemma-reach1}, all proper subterms %
of $t_{i-1}$ are reachable from either a constant or
a proper subterm of $t_0$. Since $t_i$ is a proper subterm
of $t_{i-1}$, it is reachable from either a constant or
a proper subterm of $t_0$. In either case we infer the existence of
a derivation starting from a term with height smaller
than $t_0$, and involving infinitely many rewrite steps with $R$,
which contradicts our choice of $t_0$.
Hence, the infinite derivation we picked has Property~(b).

Finally, we show that 
there are infinitely many rewrite steps at position $\lambda$.
Suppose not. Let $t_{i-1}\rightarrow_{R\cup E} t_i$
be the last rewrite step at position $\lambda$. Then, 
there is a derivation starting from
some subterm at depth $1$ of $t_i$ with infinitely many
rewrite steps with $R$.
As before, this subterm is reachable
from either a constant or a proper subterm of $t_0$.
Again, this implies the existence of an infinite derivation
that starts from a term with height smaller than $t_0$.
This contradicts the minimality of $t_0$.\qed

The idea of the decidability proof is the same as that
for Theorem~\ref{theorem-innermost}, that is, we show
that if $R/E$ is non-terminating, then it is non-terminating
from a ground flat term.

\begin{lemma}\label{lemma-main}
If $R/E$ is non-terminating (innermost non-terminating), then
there exists an (innermost) derivation starting from
a ground flat term with infinitely many rewrite steps.
\end{lemma}
\proof
Assume that
there is no infinite derivation %
starting from a constant.
We will show that there is one from a ground flat term.

Using Lemma~\ref{lemma-main-aux}, we know there is
an infinite derivation,
$t_0\rightarrow_{R\cup E} t_1 \rightarrow_{R\cup E}\ldots$,
with Properties~(a),~(b) and~(c) from Lemma~\ref{lemma-main-aux}.
All the infinitely many rewrite steps at position $\lambda$
in this derivation necessarily are done using rules of the form 
$l\rightarrow f(\alpha_1,\ldots,\alpha_m)$,
where the height of $l$ is greater than or equal to $1$.
(If not, then there will be a constant in the derivation.)
By Lemma~\ref{decreasing}, $\|t_{i-1}\|\geq\|t_i\|$ for all $i$.
Since this relation can not be indefinitely decreasing,
for some $n$ we have $\|t_n\|=\|t_{n+1}\|=\|t_{n+2}\|=\ldots$.
From the derivation $t_n\rightarrow_{R\cup E} t_{n+1}
\rightarrow_{R\cup E}\ldots$
we construct a new derivation
$t_n'\rightarrow_{R\cup E}^{0,1} t_{n+1}'\rightarrow_{R\cup E}^{0,1}
\ldots$ with also infinitely many rewrite steps as follows.
Analogously to the
proof of Lemma~\ref{lemma-flat-innermost}, 
we can deduce the existence of 
at least one constant $c$ that is a normal form.
For every $t_i$, we construct $t_i'$ to be equal to $t_i$
except for the subterms at depth $1$ that are not reachable
from constants, which are replaced by $c$. Formally,
$t_i'=t_i[c]_{j_1}\ldots [c]_{j_k}$ if $t_i|_{j_1},\ldots, t_i|_{j_k}$
are the subterms at depth $1$ in $t_i$ that are not reachable
from constants.

We show that the new derivation is correct by analyzing each
rewrite step $t_{i-1}\rightarrow_{R\cup E} t_i$ and its corresponding
step $t_{i-1}'\rightarrow_{R\cup E}^{0,1} t_i'$. 
\begin{enumerate}[(1)]
\item If $t_{i-1}\rightarrow_{R\cup E} t_i$
is done at a position inside a subterm at depth $1$ of $t_{i-1}$
that is reachable from a constant, then, the same rewrite
step can be applied on $t_{i-1}'$ to produce $t_i'$.
\item
If $t_{i-1}\rightarrow_{R\cup E} t_i$
is done at a position inside a subterm, say $t_{i-1}|_j$,
at depth $1$ of $t_{i-1}$
that is not reachable from a constant, then,
$t_i|_j$ is neither reachable from a constant.
This follows from $\|t_i|_j\|=\|t_{i-1}|_j\|\geq 1$
and the fact that, by Lemma~\ref{corollary-reach1},
if a term is reachable from a constant, then all
its subterms are. Thus, $t_{i-1}'=t_i'$ holds, and
hence, $t_{i-1}'\rightarrow^0 t_i'$ holds.
\item
If $t_{i-1}\rightarrow_{R\cup E} t_i$
is done at position $\lambda$, then, by Lemma~\ref{decreasing},
if $f(l_1,\ldots,l_m)\rightarrow r$ and $\sigma$ are the rule
and substitution applied,
then $l_k$ is a variable for 
every position $k$ such that $t_{i-1}|_k$ 
is not reachable from a constant.
We define a new substitution $\sigma'$
to be equal to $\sigma$ except for such variables $l_k$,
for which we define $\sigma'(l_k)=c$. The same rule
$f(l_1,\ldots,l_m)\rightarrow r$ applied to $t_{i-1}'$
at position $\lambda$ and with substitution $\sigma'$
produces $t_i'$.
\end{enumerate}
Since every rewrite step $t_{i-1}\rightarrow_{R\cup E} t_i$
at position $\lambda$ corresponds to a rewrite step
$t_{i-1}'\rightarrow_{R\cup E}^1 t_i'$, and there are infinitely many of
such steps, it follows that the derivation
$t_n'\rightarrow_{R\cup E}^{0,1} t_{n+1}'\rightarrow_{R\cup E}^{0,1}\ldots$
has infinitely many rewrite steps.

Note that all subterms at depth $1$ in $t_n'$
are reachable from constants. Therefore,
there exists a ground flat term $t$ with height $1$ such that
$t\rightarrow_{R\cup E}^* t_n'$ holds, and hence, there exists
an infinite derivation from a ground flat term $t$.
To finish the proof, we only need to prove that this
infinite derivation has infinitely many rewrite steps with $R$.

We proceed by contradiction
by assuming that
$t_n'\rightarrow_{R\cup E}^{0,1} t_{n+1}'\rightarrow_{R\cup
E}^{0,1}\ldots$ contains only finitely many rewrite steps
with $R$.  Hence, there exists an $N\geq n$ such that
the derivation
$t_N'\rightarrow_{R\cup E}^{0,1} t_{N+1}'\rightarrow_{R\cup
E}^{0,1}\ldots$ contains {\em{no}} steps with $R$.
Call this derivation $\pi'$.
We can observe the following properties about the
corresponding old derivation
$t_N\rightarrow_{R\cup E} t_{N+1}\rightarrow_{R\cup
E}\ldots$, which we name $\pi$:
\\
(a) All rewrite steps at position $\lambda$ in 
the derivation $\pi$ are performed with $E$:  
if there was a rewrite step 
$t_{i}\to_{R,\lambda} t_{i+1}$ in $\pi$, then
we would have had 
$t_{i}'\to_{R,\lambda}^{1} t_{i+1}'$ in $\pi'$, which 
contradicts the fact that there are no rewrite steps 
with $R$ in $\pi'$.
\\
(b) In $\pi$, there are infinitely many rewrite steps of
the form $t_{i}\to_{R,j.p}t_{i+1}$ where $t_i|_j$ is
not reachable from a constant:
we know that there are infinitely many rewrite steps with $R$
in $\pi$, but there are no rewrite steps with $R$ in $\pi'$,
and hence, all the (infinitely many) rewrite steps with $R$
in $\pi$ have to fall in Case~(2) above.

From facts~(a) and~(b), it follows that there is a subterm
$t_N|_{j}$ that is not reachable from a constant and
such that there is an infinite derivation
starting from $t_N|_j$ that uses infinitely many rewrites with $R$. 
This is because any subterm at depth $1$ in the derivation $\pi$
that is not reachable from a constant is either
(i) left unchanged by a rewrite step in $\pi$ (it may be moved to
another position at depth $1$), or
(ii) it is rewritten into a subterm at depth $1$ that is also not
reachable from a constant (because of the choice of $n$ and the fact
that $N\geq n$). A subterm that is reachable from a constant can not
be rewritten into a term that is not reachable from a constant.

As before, the subterm $t_N|_j$ is reachable
from either a constant or a proper subterm of $t_0$.
Hence, there is an infinite derivation with infinitely
many rewrite steps with $R$ starting from a constant or
a proper subterm of $t_0$, contradicting the
minimality of $t_0$.\qed %

Now, the main result follows immediately
from Lemmas~\ref{lemma-simplifying1}, \ref{lemma-simplifying2},
\ref{lemma-simplifying3}, \ref{lemma-main} and 
Theorem~\ref{theorem-c}.

\begin{thm}
Termination and innermost termination are both
decidable for rewriting with right-shallow right-linear TRS 
modulo permutative theories.\qed
\end{thm}

\section{Termination is PSPACE-hard for flat right-linear TRS}\label{sec-hardness}

\noindent The algorithms of the previous sections decide termination by 
essentially generating
all terms reachable from ground flat terms up to a height
linearly bounded by the size of TRS $R$. Thus, these algorithms
run in double exponential time, since there are that
many different reachable terms to consider. In this section
we give a lower bound for the time complexity of
these problems.

\begin{thm}
The termination and innermost termination are PSPACE-hard for
flat right-linear TRS.
\end{thm}
\proof
We reduce from the following automata intersection
problem, which is
well-known to be PSPACE-complete~\cite{Kozen77}, to non-termination:
\\
\begin{tabular}{rl}
{\bf Input}: & $n$ finite (word) automata $A_1,\ldots,A_n$.\\
{\bf Question}: & ${\mathcal L}(A_1)\cap\ldots\cap{\mathcal
L}(A_n)\not=\emptyset$?
\end{tabular}

Let $\langle Q_1,\Sigma,q_{01},F_1,\Delta_1\rangle,\ldots,
\langle Q_n,\Sigma,q_{0n},F_n,\Delta_n\rangle$ be $A_1,\ldots,A_n$,
respectively, more explicitly written. 
Here 
$Q_i$ is the set of states,
$\Sigma$ is the alphabet,
$q_{0i}$ is the initial state,
$F_i$ is the set of final states and
$\Delta_i$ is the set of transitions of the $i$-th automaton.
Without loss of generality, we assume that $\Sigma$ is $\{a,b\}$.

Our goal is to construct a TRS $R$ satisfying that
$R$ is non-terminating if and only if
${\mathcal L}(A_1)\cap\ldots\cap{\mathcal
L}(A_n)\not=\emptyset$ holds.
It is easy to check whether the empty word $\lambda$ is
accepted by all $A_i$. In the affirmative case we may generate,
as the result of our reduction, a trivially non-terminating TRS.
Thus, from now on, assume that $\lambda$ is not in
${\mathcal L}(A_1)\cap\ldots\cap{\mathcal L}(A_n)$.

The idea behind the construction of $R$ is as follows.
A word $w$, say $aba$, is encoded by terms, either
$f(a,f(b,a))$ or $f(f(a,b),a)$.
We will include rules in $R$ so that if $w\in {\mathcal L}(A_i)$,
then $c_i$ can $R$-reach every possible representation of $w$.
If 
${\mathcal L}(A_1)\cap\ldots\cap{\mathcal L}(A_n)\not=\emptyset$,
then we would like to get a nonterminating derivation
$c \rightarrow h(c_1,\ldots,c_n) \rightarrow h(t,\ldots,t) 
\rightarrow c \rightarrow \cdots$ using the rules
$c \rightarrow h(c_1,\ldots,c_n)$ and 
$h(x,\ldots,x) \rightarrow c$ in $R$.
To ensure that ``all other rules'' of $R$ are terminating,
the constant $c_i$ will not reach all terms in ${\mathcal L}(A_i)$, but
only terms of a bounded length.

Let $M$ be $|Q_1|\cdot|Q_2|\cdot\cdots\cdot|Q_n|$.
Let $N$ be $\lceil {\tt log}_2(M)\rceil$.
Formally, $R$ is defined over the following alphabet.
$$
\begin{array}{rcl}
\bar{\Sigma}   &=&\bar{\Sigma}_0\cup\bar{\Sigma}_2\cup\bar{\Sigma_n}\\
\bar{\Sigma}_0 &=&\{a,b,c,c_1,\ldots,c_n\}\cup
               \{c_{ijq\hat{q}}|i\in\{1,\ldots,n\},j\in\{0,\ldots,N\},q,\hat{q}\in Q_i\}\\
\bar{\Sigma}_2 &=&\{f\}\\
\bar{\Sigma}_n &=&\{h\}
\end{array}
$$
$R$ is defined to contain the following rules:
$$
\begin{array}{rcll}
c&\to&h(c_1, c_2,\ldots, c_n)
\\
h(x,x,\ldots,x)&\to&c
\\
c_i &\to & c_{i0q_{0i}q} & i\in\{1,\ldots,n\},q\in F_i\}
\\
c_{ijq\hat{q}} & \to & a  & i\in\{1,\ldots,n\},j\in\{0,\ldots,N\},
(qa\to\hat{q})\in \Delta_i
\\
c_{ijq\hat{q}} & \to & b  & i\in\{1,\ldots,n\},j\in\{0,\ldots,N\},
(qb\to\hat{q})\in \Delta_i
\\
c_{ijq\hat{q}} & \to & f(c_{i(j+1)q\bar{q}},c_{i(j+1)\bar{q}\hat{q}})
&
i\in\{1,\ldots,n\},j\in\{0,\ldots,N-1\},q,\bar{q},\hat{q}\in
Q_i
\end{array}
$$
Now, we prove that
$R$ is non-terminating if and only if
${\mathcal L}(A_1)\cap\ldots\cap{\mathcal
L}(A_n)\not=\emptyset$ holds.

$\Leftarrow$:
Suppose that
${\mathcal L}(A_1)\cap\ldots\cap{\mathcal L}(A_n)$ is not empty.
In this case, it is well-known that
there exists a word 
$w\in{\mathcal L}(A_1)\cap\ldots\cap{\mathcal L}(A_n)$ 
with size bounded by $M$.
Thus, there exists a term $t$ with height bounded
by $N$, with $f$ in all its internal nodes, and
whose sequence of leaves is $w$.
It is clear that $c$ reaches
$h(t,\ldots,t)$.
Moreover, by using the rule $h(x,\ldots,x)\to c$,
this term reaches $c$. Therefore, $c\rightarrow_R^+ c$. 
Hence, $R$ is nonterminating.

$\Rightarrow$: Suppose that 
${\mathcal L}(A_1)\cap\ldots\cap{\mathcal L}(A_n)$ is empty.
In order to prove termination of $R$,
it suffices to prove termination of $R$ starting from any 
right-hand side term of $R$.
Suppose $R$ does not terminate starting from the term $s$.
\\
(a) First, we observe that $R$ terminates from
all constants of $\bar{\Sigma}_0-\{c\}$ independently of
the form of $A_1,\ldots,A_n$. 
Hence, $s\not\in\bar{\Sigma}_0-\{c\}$.
\\
(b) Consider the case when $s$ is $c$.
But, the fact that ${\mathcal L}(A_1)\cap\ldots\cap{\mathcal L}(A_n)$
is empty ensures that $R$ is also terminating from $c$, and hence
$s\not\in\bar{\Sigma}_0$.
\\
(c)
If $s$ is $h(c_1,\ldots,c_n)$, then either there is a derivation
with infinitely many rewrite steps
starting from some $c_i$ or there is a derivation 
with infinitely many rewrite steps and starting
from $c$.  We argued above that none of these cases is possible.
\\
(d)
If $s$ is $f(c_{i(j+1)q\bar{q}},c_{i(j+1)\bar{q}\hat{q}})$,
then, since there is no rule with left-hand side rooted by $f$,
there is a derivation with infinitely many rewrite
steps starting from one of the arguments.
We argued above that there are no derivations
with infinitely many rewrite steps and starting from
constants.

We finish the proof by noting that the size of $R$ is 
$O(nN\sum_{i=1}^{n}(|Q_i|^3+|\Delta_i|))$,
which is polynomial in the size $\sum_{i=1}^n(|Q_i|+|\Delta_i|)$
of the automata intersection problem.
\qed

\section{Undecidability of termination for flat TRS}\label{sec-undec}

\noindent In this section, we prove undecidability of termination for 
flat TRS, and undecidability of innermost termination for 
right-flat TRS. This is done by
a reduction from the Post correspondence problem (PCP)
restricted to nonempty strings, which is defined as:\\
\begin{tabular}{rl}
{\bf Input}: & $n$ pairs of strings 
 $\langle u_1,v_1\rangle,\ldots,\langle u_n,v_n\rangle$ s.t. 
 $u_i\not=\lambda,v_i\not=\lambda$ for all $i$
\\
{\bf Question}:  &
Does there exist $k >0$ and $i_1,\ldots,i_k$ s.t.
$1\leq i_1\leq n,\ldots,1\leq i_k \leq n$ and
\\ &
$(u_{i_1}\cdots u_{i_k}=v_{i_1}\cdots v_{i_k})$ ?
\end{tabular}
Since decidability of termination for flat TRS is equivalent to
decidability of termination for shallow TRS (Lemmas~\ref{lemma-simplifying2}
and~\ref{lemma-simplifying3}), we will prove 
undecidability of termination for shallow TRS.
Since PCP is not decidable but it is semi-decidable,
and non-termination is semi-decidable for shallow
TRS (Theorem~\ref{theorem-c}), we will
reduce PCP to non-termination of shallow TRS:
a reduction to just termination is not possible.
The reduction is given in the proof of 
Theorem~\ref{theorem-shallow-undec}, but
to provide an intuition, we first illustrate it via a small example.

Consider a PCP instance
$\langle u_1,v_1\rangle,\ldots, \langle u_n,v_n\rangle$
over a signature $\Sigma$. 
The $j$'th symbol of $u_i$ and $v_i$, whenever it exists,
is denoted by $u_{i,j}$ and $v_{i,j}$ respectively.
For example, 
$\langle aa,a\rangle,
 \langle b,aba\rangle$ 
is a PCP instance over $\Sigma = \{a,b\}$.  It has a
solution $1,2,1$ since $aa\cdot b\cdot aa = a\cdot aba\cdot a$.
We want to define a rewrite system $R$ such that $R$ is non-terminating
iff there is such a solution.  
Let $n = 2$ be the number of rules in the PCP instance
and let $L = 3$ be the maximum size of a string in the PCP instance.
We define $R$ over a signature $\Sigma'$ where
where
\begin{eqnarray}
 \Sigma' & := & \Sigma'_0 \cup \Sigma'_1 \cup \Sigma'_2\cup\Sigma'_6\cup\Sigma'_{8}
\nonumber
\\
 \Sigma'_0 & := & \{U,U',V,V',P,P',P'',A,A',A''\}
\nonumber
\\
 \Sigma'_1 & := & \{a,b\} \cup \{U_{i,j},V_{i,j},P_{i,j}\mid i\in\{1,\ldots,n\}, j\in\{1,\ldots,L\}\},
\nonumber
\\ 
 \Sigma'_2 & := & \{f_1\},  \quad 
 \Sigma'_6 \;\; := \;\;\{f_3\}, \quad 
 \Sigma'_8 \;\; := \;\;\{f_2\}
\label{eqn-sigma}
\end{eqnarray}
A potential solution, say $i_1,\ldots,i_k$, to the PCP instance is encoded
by a pair of terms $(s_{uu},s_{vv})$ where
\begin{eqnarray}
 s_{uu} & := & U_{i_1,1}\ldots U_{i_1,L}\ldots U_{i_k,1}\ldots U_{i_k,L}(U)
\nonumber
\\
 s_{vv} & := & V_{i_1,1}\ldots V_{i_1,L}\ldots V_{i_k,1}\ldots V_{i_k,L}(V)
\label{eqn-suu-svv}
\end{eqnarray}
Concretely, the solution $1,2,1$ is encoded by the pair
\begin{eqnarray*}
 s_{uu} & := & U_{1,1}U_{1,2}U_{1,3}U_{2,1}U_{2,2}U_{2,3}U_{1,1}U_{1,2}U_{1,3}(U)
\\
 s_{vv} & := & V_{1,1}V_{1,2}V_{1,3}V_{2,1}V_{2,2}V_{2,3}V_{1,1}V_{1,2}V_{1,3}(V)
\end{eqnarray*}
Here the notation 
$U_{1,1}U_{1,2}U_{1,3}(U)$ serves as a shorthand for the term
$U_{1,1}(U_{1,2}(U_{1,3}(U)))$. 
This convention allows us to view strings as (parts of) terms.
We need to construct a rewrite system $R$ whose non-termination 
implies that $s_{uu},s_{vv}$ indeed correspond
to a solution of the PCP.  Hence, we need to check that
\\
(1) $s_{uu}$ and $s_{vv}$ are of the form above,
\\
(2) the indices sequence $i_1,\ldots,i_k$ 
in $s_{uu}$ and $s_{vv}$ are the same, and
\\
(3) the words $u_{i_1}\ldots u_{i_k}$ and $v_{i_1}\ldots v_{i_k}$ are the same.

To check~(1), we introduce the following rules in $R$:
\begin{eqnarray}
R_U & := & \{ U_{i,1}U_{i,2}\cdots U_{i,L}(U) \rightarrow U', \;
	      U_{i,1}U_{i,2}\cdots U_{i,L}(U') \rightarrow U' \mid i\in\{1,\ldots,n\}\}
\nonumber
\\
R_V & := & \{ V_{i,1}V_{i,2}\cdots V_{i,L}(V) \rightarrow V', \; 
	      V_{i,1}V_{i,2}\cdots V_{i,L}(V') \rightarrow V' \mid i\in\{1,\ldots,n\}\}
\label{eqn-ru}
\end{eqnarray}
We note that $s_{uu} \rightarrow_{R_U}^* U'$ and
$s_{vv} \rightarrow_{R_V}^* V'$.
Hence we can check~(1) by checking if 
$s_{uu} \rightarrow_R^* U'$
and
$s_{vv} \rightarrow_R^* V'$.
But this does not still check that the sequence $i_1,\ldots,i_k$
(sequence $1,2,1$ in the example) used
in $s_{uu}$ is the same as the one used in $s_{vv}$.

To check~(2), we make $U_{i,j}(x)$ and $V_{i,j}(x)$ rewrite to
$P_{i,j}(x)$.  Hence, we introduce the following rules in $R$:
\begin{eqnarray}
R_{2P} & := & \{ U_{i,j}(x) \rightarrow P_{i,j}(x), V_{i,j}(x) \rightarrow P_{i,j}(x) \mid i\in\{1,\ldots,n\}, j\in\{1,\ldots,L\}\}
\nonumber
\\
R_{P'} & := & \{ P_{i,1}P_{i,2}\cdots P_{i,L}(P') \rightarrow P' \mid i\in\{1,\ldots,n\}\}
\nonumber
\\
R_{P''} & := & \{ P_{i,1}P_{i,2}\cdots P_{i,L}(P'') \rightarrow P'' \mid i\in\{1,\ldots,n\}\}
\nonumber
\\
R_{UP} & := & \{ U \rightarrow P,  V \rightarrow P, P \rightarrow P', P \rightarrow P''\}
\label{eqn-rp}
\end{eqnarray}
Now, using these new rules,
we note that $s_{uu}$ and $s_{vv}$ are joinable if
they use the same sequence of indices $i_1,\ldots,i_k$.
In fact, both $s_{uu}$ and $s_{vv}$ rewrite to the term
$s_{pp}$, where
$$
 s_{pp} := P_{i_1,1}\cdots P_{i_1,L}\cdots P_{i_k,1}\cdots P_{i_k,L}(P)
$$
Moreover, using $R_{P''}\cup R_{UP}$, 
$s_{pp}$ can rewrite to either $P'$ or $P''$.
Thus, we can check~(2) by checking for the joinability of 
$s_{uu}$ and $s_{vv}$ to a term that can reach both $P'$ and $P''$.

Finally, to check~(3), we introduce the following rules in $R$:
\begin{eqnarray}
R_{UA} & := & \{ U \rightarrow A, V \rightarrow A, A \rightarrow A', A\rightarrow A''\}
\nonumber
\\
R_\alpha & := &  \{ \alpha(A')\rightarrow A', \alpha(A'')\rightarrow A'' \mid \alpha\in\Sigma \}
\nonumber
\\
R_w & := & \{U_{i,j}(x)\rightarrow u_{i,j}(x) \mid 1\leq j\leq |u_i|\}
\cup \{U_{i,j}(x)\rightarrow x \mid j>|u_i|\}
\nonumber
\\ & & 
\cup \{V_{i,j}(x)\rightarrow v_{i,j}(x) \mid 1\leq j\leq |v_i|\}
\cup \{V_{i,j}(x)\rightarrow x \mid j>|v_i|\}
\label{eqn-ra}
\end{eqnarray}
Using these rules,
$s_{uu}$ and $s_{vv}$ can both rewrite to a common term 
($w(A)$) if
the strings 
$u_{i_1}\ldots u_{i_k}$ 
and
$v_{i_1}\ldots v_{i_k}$ are equal (to $w$).
In our example, $w := aabaa$.
Moreover, the common reachable term ($w(A)$) can then rewrite to either $A'$ or $A''$.
Hence, we can check~(3) by checking for joinability of $s_{uu}$ and
$s_{vv}$ to a term that can reach both $A'$ and $A''$.

We can put everything together by introducing three more rules in $R$:
\begin{eqnarray}
R_f & := & \{f_1(x,y)\rightarrow f_2(x,y,x,y,x,y,x,y), 
	f_2(x,y,z,z,z',z',U',V')\rightarrow f_3(x,y,z,z,z',z'), 
\nonumber
	\\ & & 
	f_3(x,y,A',A'',P',P'')\rightarrow f_1(x,y) \}
\label{eqn-rf}
\end{eqnarray}
If $s_{uu},s_{vv}$ is generated from a solution of the PCP instance, then
we can immediately get a nonterminating derivation using $R$:
\begin{alignat}{2} %
f_1(s_{uu},s_{vv}) & \rightarrow_{R_f} &\;&
	f_2(s_{uu},s_{vv},s_{uu},s_{vv},s_{uu},s_{vv},s_{uu},s_{vv})
\nonumber
\\ & \rightarrow_{R_U,R_V}^* &&
	f_2(s_{uu},s_{vv},s_{uu},s_{vv},s_{uu},s_{vv},U',V')
\nonumber
\\ & \rightarrow_{R_{2P},R_{UP}}^* &&
	f_2(s_{uu},s_{vv},s_{uu},s_{vv},s_{pp},s_{pp},U',V')
\nonumber
\\ & \rightarrow_{R_{w},R_{UA}}^* &&
	f_2(s_{uu},s_{vv},w(A),w(A),s_{pp},s_{pp},U',V')
\nonumber
\\ & \rightarrow_{R_{f}} &&
	f_3(s_{uu},s_{vv},w(A),w(A),s_{pp},s_{pp})
\nonumber
\\ & \rightarrow_{R_{P'},R_{P''},R_{UP}}^* &&
	f_3(s_{uu},s_{vv},w(A),w(A),P',P'')
\nonumber
\\ & \rightarrow_{R_{\alpha},R_{UA}}^* &&
	f_3(s_{uu},s_{vv},A',A'',P',P'')
\nonumber
\\ & \rightarrow_{R_{f}} &&
	f_1(s_{uu},s_{vv})
\label{eqn-infinite-derivation}
\end{alignat}
The following theorem formally describes and proves this reduction.

\begin{thm}\label{theorem-shallow-undec}
Termination of shallow TRS is undecidable.
\end{thm}
\proof
Consider an instance 
$\langle u_1,v_1\rangle,\ldots,\langle u_n,v_n\rangle$ of the
restricted PCP, that is,
$u_i, v_i$ are nonempty strings over alphabet $\Sigma$.
We construct a shallow TRS $R$ such that
this PCP instance  has a solution iff $R$ is non-terminating.

Let $L={\tt Max}(|u_1|,\ldots,|u_n|,|v_1|,\ldots,|v_n|)$.
We construct $R$ over a signature $\Sigma'$, where $\Sigma'$ is defined in
Equation~\ref{eqn-sigma}.
The TRS $R$ is defined as follows:
\begin{eqnarray*}
R & := & R_U \cup R_V \cup R_{2P} \cup R_{P'}\cup R_{P''}\cup R_{UP}\cup \cup R_{\alpha}\cup R_{w} \cup R_{UA} \cup R_f
\end{eqnarray*}
where 
$R_U,R_V$ are defined in Equation~\ref{eqn-ru},
$R_{2P},R_{P'},R_{P''},R_{UP}$ are defined in Equation~\ref{eqn-rp},
$R_{\alpha},R_{w},R_{UA}$ are defined in Equation~\ref{eqn-ra}
and
$R_f$ is defined in Equation~\ref{eqn-rf}.

\ignore{
In order to see how this works, we will call
$U_i(x)$ to the term $U_{i,1}\ldots U_{i,L}(x)$,
$V_i(x)$ to the term $V_{i,1}\ldots V_{i,L}(x)$,
$P_i(x)$ to the term $P_{i,1}\ldots P_{i,L}(x)$,

We want to see that, if there is a solution of the original PCP
problem, then the resulting TRS is not terminating.
\endignore}

$\Rightarrow$:
We first show that if the PCP instance has a solution, then
$R$ is non-terminating.
Let $i_1,\ldots,i_k$
be a solution of the PCP instance,
i.e.\ $w=u_{i_1}\ldots u_{i_k}=v_{i_1}\ldots v_{i_k}$ holds.
Then, we have the infinite derivation, shown in Equation~\ref{eqn-infinite-derivation},
starting from the ground term $s_1 := f_1(s_{uu}, s_{vv})$, where
$s_{uu}$ and $s_{vv}$ are defined in Equation~\ref{eqn-suu-svv}.

$\Leftarrow$:
Suppose $R$ does not terminate. We need to show that the PCP 
instance has a solution.
To this end we define the concept of {\em $UV$-variant}.
We say that a term $s$ is a $UV$-variant of a term $t$, if
$t$ can be obtained from $s$ by applying several rewrite
steps using rules from the subset
$\{U_{i,j}(x)\rightarrow x: i=1\ldots n,j>|u_i|\}
\cup \{V_{i,j}(x)\rightarrow x: i=1\ldots n,j>|v_i|\}$ of $R_w$.
Note that, since none of $u_i$ or $v_i$ is $\lambda$ in 
the original PCP instance,
$s$ and $t$ have the same number
of occurrences of symbols of $\{U_{i,1}:i=1\ldots n\}\cup
\{V_{i,1}:i=1\ldots n\}$.

Now, note that since all rules in $R$ are height-preserving or 
height-decreasing,
there is a derivation with infinitely many rewrite steps at the top.
We pick such a derivation, but with minimal height for
the initial term $t$. 
Then, the root symbol of $t$ has to be one of the
$f_i$'s: otherwise, only a finite number of rewrite
steps can be done at the top
and preserving the height.
Therefore, we have a derivation of the form
$f_1(\ldots)\rightarrow^* f_2(\ldots)\rightarrow^* f_3(\ldots)\rightarrow^*
f_1(\ldots)\rightarrow^*\ldots$ with infinitely many rewrite
steps at the top. We can assume that we start
with a term of the form $f_1(u,v)$. By observing the $R_f$ rules,
one can deduce that $u$ and $v$ reach $A',A'',P',P''$, and that
$u$ reaches $U'$ and that $v$ reaches $V'$.
This is possible only if the terms $u$ and $v$ are
$UV$-variants of terms of the form
\begin{eqnarray*}
 s_{uu} & := & U_{i_1,1}\ldots U_{i_1,L}\ldots U_{i_k,1}\ldots U_{i_k,L}(U)
\\
 s_{vv} & := & V_{j_1,1}\ldots V_{j_1,L}\ldots V_{j_{k'},1}\ldots V_{j_{k'},L}(V)
\end{eqnarray*}
where $k,k'\geq 1$.
But, moreover, these terms have to be joinable to a term of the form
$P_{i_1,1}\ldots P_{i_1,L}\ldots P_{i_k,1}\ldots P_{i_k,L}$ $(P)$,
and also of the form $P_{j_1,1}\ldots P_{j_1,L}\ldots
P_{j_{k'},1}\ldots P_{j_{k'},L}(P)$. 
(Note here that since $u_i,v_i$ are not $\lambda$, terms like
$U_{i,1}\ldots U_{i,L}(x)$ can not rewrite to $x$ and hence
the indices $i_1,\ldots,i_k,j_1,\ldots,j_{k'}$ will be preserved
in any joinability proof.)
Hence, $k=k'$ and $i_r=j_r$ for all $r$ in $\{1,\ldots,k\}$. But moreover,
$u$ and $v$ have to be joinable to a term of the form
$u_{i_1}\ldots u_{i_k}(A)=v_{i_1}\ldots v_{i_k}(A)$. Hence,
$u_{i_1}\ldots u_{i_k}=v_{i_1}\ldots v_{i_k}$ and there is a solution
of the original PCP.  \qed

\noindent
{\em Remark:}
It is important to keep $P'$ and $P''$ (and $A'$ and $A''$) as two different
constants in the above proof.  If we reuse $P$ in place of $P'$ and $P''$
(respectively, $A$ in place of $A'$ and $A''$),
then terms that satisfy Check~(1), but do not satisfy Check~(2) 
(respectively, Check~(3)), such as, 
$s_{uu} := U_{1,1}U_{1,2}U_{1,3}(U)$ and $s_{vv} := V_{2,1}V_{2,2}V_{2,3}(V)$, 
which do {\em{not}} correspond to a solution of the PCP,
would generate infinite derivations  starting from $f(s_{uu},s_{vv})$.

Combining Theorem~\ref{theorem-shallow-undec} with 
Lemmas~\ref{lemma-simplifying2}
and~\ref{lemma-simplifying3}, we have the following result.

\begin{thm}\label{theorem-flat-undec}
Termination of flat TRS is undecidable.\qed
\end{thm}

For the case of innermost rewriting, we have
seen that termination is decidable for flat TRS.
However, in the innermost case we have the following result.

\begin{thm}\label{thm-inner-right-flat-undecidable}
Innermost termination
of right-flat TRS is undecidable.
\end{thm}
\proof
Given an instance 
$\langle u_i, v_i\rangle,\ldots,\langle u_n,v_n\rangle$
of Post correspondence problem,
we generate the TRS
$R = \{f(x) \rightarrow g(x,x,x),  \;
 g(x, u_i(y), v_i(z))\rightarrow h(x,y,z),  \;
 h(x, u_i(y), v_i(z))\rightarrow h(x,y,z),  \;
 h(x, \epsilon, \epsilon) \rightarrow f(x) \mid 1\leq i\leq n\}$.
Here $\epsilon$ is a constant representing the empty string.
Note that $R$ is right-flat.
It is easy to see that the PCP instance has a solution
iff $R$ is innermost non-terminating.
\qed

\noindent
{\em Remark:}
A reduction similar to the one in the proof of 
Theorem~\ref{thm-inner-right-flat-undecidable}
was given in Definition 5.3.6 of~\cite{TeReSe} for
showing undecidability of termination for 
(general) term rewriting systems.

\section{Conclusions}

\noindent We showed that termination and innermost termination 
of right-shallow right-linear term rewriting systems
is decidable.
This result also holds when we consider rewriting modulo permutative
theories.
We also showed that innermost termination of flat TRSs
is decidable, whereas termination of flat TRSs is undecidable.
For the decidable problems,
the complexity of the given algorithms
is doubly exponential, whereas we have also
provided a PSPACE-hardness lower bound.
It is unclear whether both upper and lower bounds can be improved
in some way. As further work it would be interesting
to fix the exact complexity of these problems, but also to
consider other classes of TRS,  for example, classes defined 
by imposing syntactic restrictions not on the original TRS,
but on the dependency pairs of the TRS~\cite{Sakai06,Sakai}.


\begin{thebibliography}{BKdV03}

\bibitem[BKdV03]{TeReSe}
M.~Bezem, J.~W. Klop, and R.~de~Vrijer, editors.
\newblock {\em Term Rewriting Systems}.
\newblock Cambridge Tracts in Theoretical Computer Science 55. Cambridge
  University Press, 2003.

\bibitem[BN98]{Allthat}
F.~Baader and T.~Nipkow.
\newblock {\em Term Rewriting and All That}.
\newblock Cambridge University Press, New York, 1998.

\bibitem[Der81]{Dershowitz81:ICALP}
N.~Dershowitz.
\newblock Termination of linear rewriting systems.
\newblock In {\em Proc. 8th Colloquium on Automata, Languages and Programming,
  ICALP}, volume 115 of {\em LNCS}, pages 448--458, 1981.

\bibitem[GHT07]{DBLP:conf/rta/GodoyHT07}
G.~Godoy, E.~Huntingford, and A.~Tiwari.
\newblock Termination of rewriting with right-flat rules.
\newblock In {\em Proc. 18th Intl. Conf. on Rewriting Techniques and
  Applications, RTA}, volume 4533 of {\em LNCS}, pages 200--213, 2007.

\bibitem[GT04]{GodoyTiwari04:IJCAR}
G.~Godoy and A.~Tiwari.
\newblock Deciding fundamental properties of right-(ground or variable) rewrite
  systems by rewrite closure.
\newblock In {\em Proc. Intl. Joint Conf. on Automated Deduction, IJCAR},
  volume 3097 of {\em LNAI}, pages 91--106. Springer, July 2004.

\bibitem[GT05]{GodoyTiwari05:CADE}
G.~Godoy and A.~Tiwari.
\newblock Termination of rewrite systems with shallow right-linear, collapsing,
  and right-ground rules.
\newblock In {\em Proc. 20th Intl. Conf. on Automated Deduction, CADE}, volume
  3632 of {\em LNCS}, pages 164--176. Springer, July 2005.

\bibitem[HL78]{HuetLankford78}
G.~Huet and D.~S. Lankford.
\newblock {\em On the uniform halting problem for term rewriting systems}.
\newblock INRIA, Le Chesnay, France, 1978.
\newblock Technical Report 283.

\bibitem[Koz77]{Kozen77}
D.~Kozen.
\newblock Lower bounds for natural proof systems.
\newblock In {\em Proc. 18th Symp. on the Foundations of Computer Science},
  pages 254--266, 1977.

\bibitem[Pla93]{Plaisted93:RTA}
D.~A. Plaisted.
\newblock Polynomial time termination and constraint satisfaction tests.
\newblock In {\em Proc. 5th Intl. Conf. on Rewriting Techniques and
  Applications, RTA}, volume 690 of {\em LNCS}, pages 405--420, 1993.

\bibitem[TKS00]{TakaiKajiSeki00:RTA}
T.~Takai, Y.~Kaji, and H.~Seki.
\newblock Right-linear finite path overlapping term rewriting systems
  effectively preserve recognizability.
\newblock In {\em Proc. 11th Intl. Conf. on Rewriting Techniques and
  Applications, RTA}, volume 1833 of {\em LNCS}, pages 246--260, 2000.

\bibitem[Toy87]{Toy87b}
Y.~Toyama.
\newblock Counterexamples to termination for the direct sum of term rewriting
  systems.
\newblock {\em Information Processing Letters}, 25:141--143, 1987.

\bibitem[USS10]{Sakai}
K.~Uchiyama, M.~Sakai, and T.~Sakabe.
\newblock Decidability of termination and innermost termination for term
  rewriting systems with right-shallow dependency pairs.
\newblock {\em {IEICE} Trans. on Information and Systems}, E93-D(5):953--962,
  2010.

\bibitem[WS06]{Sakai06}
Y.~Wang and M.~Sakai.
\newblock Decidability of termination for semi-constructor trss, left-linear
  shallow trss and related systems.
\newblock In {\em Proc. 17th Intl. Conf. on Rewriting Techniques and
  Applications, RTA}, volume 4098 of {\em LNCS}, pages 343--356. Springer,
  2006.

\end{thebibliography}
\end{document}